\documentstyle[aps,twocolumn,epsf,tighten]{revtex}

\begin{document}
\pagestyle{empty}                                      
\draft
\vfill
%
\title{Glueball States in a Constituent Gluon Model}
\vfill
\author{$^{1}$Wei-Shu Hou, $^{1}$Ching-Shan Luo and $^{2}$Gwo-Guang Wong}
\address{
\rm $^{1}$Department of Physics, National Taiwan University,
Taipei, Taiwan 10764, ROC\\ 
and\\
\rm $^{2}$Department of International Trade, Fushin Institute of Technology,
Toucheng, Ilan, Taiwan 26141, ROC
}
\date{\today}

\vfill
\maketitle
\begin{abstract}
In a model with dynamical gluon mass,
we investigate the bound states of 
two and three gluons via a Schr\"odinger equation. 
The short distance potential is approximated by one-gluon-exchange 
while the long distance part is assumed to be of a breakable string. 
We estimate the masses and in particular the {\it sizes} of 
low-lying bound states with no orbital angular momentum. 
By considering quantum-mechanical smearing of the gluon fields
and normalizing to lattice results on $M_{0^{++}}$ and $M_{2^{++}}$,
we find that the $0^{++}$ glueball is rather small in size
compared with the others.
The fitted gluon mass is of order 600 to 700 MeV, which is reasonable.
The $0^{-+}$, $1^{--}$ and $3^{--}$ three gluon glueball states 
are nearly degenerate, and their mass ratio with $2^{++}$
is largely independent of all parameters
and consistent with lattice calculations.
We estimate the mass of the $1^{--}$ glueball to be 
around $3.1-3.7$~GeV, 
which is close to the mass of $J/\psi$ and $\psi^\prime$.
\end{abstract}
\pacs{PACS numbers:}

%
\pagestyle{plain}

\section{Introduction}

During the last 20 years there has been much effort in trying to obtain 
a nonperturbative form for the gluon propagator. 
Perhaps one of the most interesting result is that
the gluon may have a dynamically generated mass~\cite{cornwall}. 
The existence of a mass scale, or the absence of a pole at $k^2=0$, 
is natural if one assumes that gluons do not propagate to infinity, 
{\it i.e.} these propagators describe confined gluons. 
The concept of massive gluon has been widely used in 
independent field theoretic studies, and
examples about the consequences of massive gluons can be found 
in the literature~\cite{parisi,halzen,field,Consoli,Papavassiliou,Natale}.
The infrared behavior of the gluon propagator
has also been studied numerically~\cite{Bernard,amundsen},
and recent lattice computations give strong evidence for
an infrared finite gluon propagator~\cite{Leinweber}.

The gluon self-coupling in quantum chromodynamics (QCD) 
implies the existence of bound states of gauge fields known as glueballs. 
Numerous technical difficulties have so far
hampered our understanding of their properties in experiments, 
largely because glueball states 
can mix strongly with nearby $q \bar q$ resonances.
However, recent experimental and lattice studies of 
$0^{++}$, $2^{++}$ and $0^{-+}$ glueballs seem to be converging.
All simulations agree that the lightest scalar glueball mass is 
in the range of $1500-1750$ MeV, while 
the tensor and pseudoscalar masses are 
in the range of $2000-2400$~MeV~\cite{Morningstar,Weingarten,Liu}.
It has been suggested~\cite{Bugg} that
improved action lattice predictions~\cite{Morningstar} 
agree well with the mass ratios of
$f_0(1500)$, $\eta (2190)$, and $f_2(1980)$,
which have exotic features that
make them natural candidates for glueballs.
All these states are seen in $p \bar p$
annihilation, central production in $pp$ collisions, or
$J/\psi\rightarrow\gamma+X$ transitions~\cite{Anisovich,Bugg2,Barberis}.

In this paper,
we reopen the case of the potential model with massive constituent gluons,
namely, the model of Cornwall and Soni~\cite{cands,Hou}. 
It is not our purpose to pursue the detailed theoretical bearings for gluon mass. 
But rather, we wish to explore how recent results in this field affect 
the potential model, and how the potential model can provide more 
insight on glueball properties.
While the potential model has its limitations, 
it gives bound state solutions that have the advantage of 
providing information such as the size of glueballs, 
an aspect which is rarely \cite{deForcrandLiu} mentioned in the literature.
In Sect. 2, we give the details of model description for 
low-lying bound states of 2-gluon and 3-gluon glueballs. 
We then use the variational method 
to estimate their masses and sizes in Sect.~3,
where some smearing of gluon field is developed for the $0^{++}$ case.
Finally, we analyze our results and 
make some conclusions in Sect. 4.

\section{Model Description}

Although the dynamically generated mass $m^2(q^2)$ should be scale 
dependent, phenomenologically we shall treat $m^2(q^2)$ as constant for 
simplicity.
The Lagrangian for the massive vector fields $A_\mu^a$ is 
\begin{equation}
{\cal L}=-{1 \over 4}F_{\mu\nu}^aF^{a\mu\nu}
+{1 \over 2}m^2A_\mu^aA^{a\mu},
\end{equation}
where $m$ is the effective gluon mass defined by Cornwall~\cite{cornwall}
and others, and 
\begin{equation}
F^a_{\mu\nu}=\partial_\mu A^a_\nu-\partial_\nu A^a_\mu
+gf^{abc}A^b_\mu A^c_\nu.
\end{equation}
The propagator for $A^a_\mu$ is
\begin{equation}
D^{ab}_{\mu\nu}(k)={-i\delta^{ab}(g^{\mu\nu}
-{k^\mu k^\nu \over m^2}) \over k^2-m^2+i\epsilon}.
\end{equation}
In the nonrelativisic limit, we expand the massive gluon 
momentum and polarization vector to order $|{\bf k}|^2$,
\begin{eqnarray} 
k^\mu &\cong& \left(m+{{\bf k}^2 \over 2m}\ ,\ {\bf k}\right),
\\ \label{m40}
\epsilon^\mu(k) &\cong& \left({{\bf k}\cdot{\bf e} \over m}\ ,\
{\bf e}+{{\bf k}\cdot{\bf e} \over 2m^2}{\bf k}\right).
\label{m30}
\end{eqnarray} 

\begin{figure}[t!]  
\centerline{{\epsfxsize3.2 in \epsffile{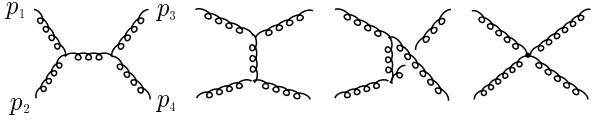}}}  
\smallskip  
\caption{$gg \rightarrow gg$ scattering.}  
\end{figure}

Let us consider the system of a 2-gluon glueball. 
There are four diagrams 
contributing at tree level in the gluon-gluon
interaction shown as in Fig. 1, corresponding to 
$s$-, $t$-, $u$-channel gluon exchange
and the seagull. 
In the non-relativistic limit, it turns out that all matrix 
elements are zeroth order in momentum except the contribution 
from $s$-channel, which can be removed both because of color and its 
second order nature in momentum. 
The short distance potential can be extracted 
from the tree-level Feynman amplitude of Fig. 1,
\begin{equation}
V({\bf r})=\int {d^3q \over (2\pi)^3}
{ie^{i{\bf q}\cdot{\bf r}} \over 4\sqrt{E_{1f}E_{2f}E_{1i}E_{2i}}}
i{\cal M}_{fi},
\label{m34}
\end{equation}
where ${\bf q}$ is the momentum transfer of the system. 

When extracting the potential, 
the ``exchange" or ``symmetric" diagrams
are automatically taken care of by 
the properly symmetrized wave function
for identical particle systems. 
Hence the relavent
contributions are
\begin{equation}
i{\cal M}_{fi}^t=-ig^2f^{ace}f^{bde}\langle 3|J_\rho|1 \rangle 
{1 \over t-m^2}\langle 4|J^\rho|2 \rangle,
\label{m44}   
\end{equation}
where
\[
\langle 3|J_\rho|1 \rangle=\epsilon_3^*\cdot\epsilon_1\,
(p_1+p_3)_\rho-2\epsilon_{3\rho}^*\, p_3\cdot\epsilon_1
-2\epsilon_{1\rho}\, p_1\cdot\epsilon_3^*,
\]
and 
\begin{eqnarray}
i{{\cal M}_{fi}^{s.g.}} & = &-ig^2\Bigl[ f^{abe}f^{cde}
(\epsilon_1\cdot\epsilon_3^*\, \epsilon_2\cdot\epsilon_4^*) \nonumber \\ & & 
+f^{ace}f^{bde}
(\epsilon_1\cdot\epsilon_2\, \epsilon_3^*\cdot\epsilon_4^*
-\epsilon_1\cdot\epsilon_4^*\,\epsilon_2\cdot\epsilon_3^*)\Bigr].
\label{m36}
\end{eqnarray}

We define the spin operator ${\bf S}=(S^1,S^2,S^3)$ as
%
$(S^k)_{ij}=-i\epsilon^{ijk}$,
which satisfy
\begin{equation}
[S^i,S^j]=i\epsilon^{ijk}S^k,
\label{m39}
\end{equation}
{\it i.e.} $S^1$, $S^2$ and $S^3$ are 
${\rm SU}(2)$ group generators as desired.
The spin operator can then be extracted via the relation
\begin{equation}
A_iB_j=\Bigl[ {\bf A}\cdot{\bf B}-({\bf S}\cdot{\bf B})
({\bf S}\cdot{\bf A})\Bigr]_{ij},
\end{equation}
and placing the polarization vectors into the wavefunction.
We calculate the matrix elements in the center of mass frame, namely,
${\bf p}_i\equiv {\bf p}_1=-{\bf p}_2$, 
${\bf p}_f\equiv {\bf p}_3=-{\bf p}_4$, and
the momentum transfer
${\bf q}\equiv {\bf p}_3-{\bf p}_1={\bf p}_f-{\bf p}_i$.
After some simplifications, we obtain
\begin{eqnarray}
i{\cal M}_{fi}^t &=& {ig^2f^{ace}f^{bde} \over {\bf q}^2+m^2}
\Bigl[4m^2+3{\bf q}^2-2{\bf S}^2{\bf q}^2+2({\bf S}\cdot{\bf q})^2
 \nonumber \\
&& + 6i{\bf S}\cdot({\bf q}\times{\bf p}_i)\Bigr], \\
i{\cal M}^{s.g.}_{fi} &=& ig^2\left[ f^{abe}f^{cde}-f^{ace}f^{bde}
\Bigl({1 \over 2}{\bf S}^2-2\Bigr)\right],
\end{eqnarray}
where ${\bf S}\equiv{\bf S}_1+{\bf S}_2$ is 
the total spin of the 2-gluon glueball. 
As stated, we have transfered the spin content of gluons 
to the wavefunction, and hence view $i{\cal M}_{fi}$ as 
an operator acting on spin space. 
Note also that 
${\bf S_1}$ acts on ${\bf e_1}$ and ${\bf e_3}$ while 
${\bf S_2}$ acts on ${\bf e_2}$ and ${\bf e_4}$.  

Using Eq. (\ref{m34}), we arrive at 
the short distance potential
\begin{eqnarray}
V_{\rm sd}(r) &=&-{g^2f^{ace}f^{bde} \over 4\pi}\Biggl\{
\biggl[{1 \over 4}+{1 \over 3}{\bf S}^2
 +{3 \over 2m^2}({\bf L}\cdot{\bf S}) 
{1 \over r}{\partial \over \partial r}\Biggr. \biggr.
 \nonumber \\
&& -{1 \over 2m^2} \biggl.
\Bigl(({\bf S}\cdot\nabla)^2-{1 \over 3}{\bf S}^2\nabla^2\Bigr)\biggr]
{e^{-mr} \over r}\label{m69}\\
&& +\Bigl(1-{5 \over 6}{\bf S}^2\Bigr)
\Biggl. {\pi \over m^2}\delta^3({\bf r})\Biggr\} 
+{g^2f^{abe}f^{cde} \over 4\pi}{\pi \over m^2}\delta^3({\bf r}),
\nonumber 
\end{eqnarray}
where ${\rm sd}$ 
stands for ``short distance.''

The gluon-gluon interaction potential 
$V_{\rm sd}$ in Eq. (\ref{m69}) is the Fourier transform of 
the tree level second order scattering diagrams of Fig. 1, 
but it cannot account for gluon confinement 
since it is of short distance nature.
We must add a term to take into
account such long distance effects. We add to $V_{\rm sd}$ a string 
potential $V_{\rm str}$ which is assumed to be spin independent,
\begin{equation}
V_{\rm str}=2m(1-e^{-\beta mr}),
\end{equation}
where $\beta$ is related to the adjoint string tension $K_A$ via
\begin{equation}
\beta={K_A \over 2m^2}.
\end{equation}
In the potential $V_{\rm str}$, the color screening of gluons is brought 
about by a breakable string, that is, the adjoint string breaks when 
sufficient energy has been stored in it to materialize a gluon pair. 
This form of the string potential simulates the 
intergluonic potential as seen in lattice calculations~\cite{Bernard}.

We thus get the gluon-gluon potential
relevant to two-gluon glueballs
\begin{eqnarray}
V_{2g}(r) &=&-\lambda\Biggl\{
\biggl[{1 \over 4}+{1 \over 3}{\bf S}^2 
+{3 \over 2m^2}({\bf L}\cdot{\bf S}) 
{1 \over r}{\partial \over \partial r}\biggr. \Biggr.
 \nonumber \\
&& -{1 \over 2m^2} \biggl. \Bigl(({\bf S}\cdot\nabla)^2
-{1 \over 3}{\bf S}^2\nabla^2\Bigr)\biggr]
{e^{-mr} \over r}  \nonumber \\
&& +\Biggl. \Bigl(1-{5 \over 6}{\bf S}^2\Bigr)
    {\pi \over m^2}\delta^3({\bf r})\Biggr\} 
+2m(1-e^{-\beta mr}), \label{m82}
\end{eqnarray} 
where $\lambda$ is defined as
\begin{equation}
\lambda\equiv{3g^2 \over 4\pi},
\label{lam} 
\end{equation}
and is related to the strong coupling strength of the process.
Note that the $f$-type constant in the last term of Eq. (\ref{m69}) 
does not contribute 
when contracted with the normalized color wavefunction 
\begin{equation}
\psi_{\rm color} (a,b)={1 \over \sqrt 8}\delta^{ab}.
\end{equation}

We are left with three parameters: effective gluon mass $m$, 
string breaking parameter $\beta$ 
and adjoint strong coupling constant $\lambda$. 
We take~\cite{Bernard} the conservative range $\beta \sim (1\pm 0.7)$,
while $\lambda$ is determined by~\cite{cands} 
\begin{equation}
\lambda={N \over 4\pi}\left[{11N \over 48\pi^2}
\ln\Bigl({4m^2 \over \Lambda^2}\Bigr)\right]^{-1},
\end{equation}
for ${\rm SU}(N)$ group. 
For $N = 3$, taking $m \sim$ 600 MeV, 
$\Lambda \sim$ 350 MeV, one gets $\lambda \sim 1.4$. For $\Lambda$
varying from $250-400$ MeV, $\lambda$ varies in the range $1.1-1.6$.

For the case of 3-gluon glueballs, 
we assume that the constituent gluons interact
pair-wise~\cite{Hou}. Thus 
\begin{equation}
V_{3g}=\sum_{i<j}\Bigl[V_{\rm sd}(r_{ij})
 +{1\over 2}V_{\rm str}(r_{ij})\Bigr]. 
\end{equation}
We note that the contribution from $s$-channel gluon exchange 
can still be ignored
because of its second order nature in momentum. 
The factor of one-half for the string potential is because one needs 
to pull three (and not six) gluons from the vacuum to screen the three
gluons that are originally there in the glueball. 

For the low-lying bound states with relative angular
momentum $l_{ij}=0$ for each pair of gluons in the 3-gluon system,
the normalized color singlet wavefunction is either symmetric 
or antisymmetric: 
\begin{equation}
\psi_{\rm color}^{d-{\rm type}}(a,b,c)=\sqrt {3\over 40}d^{a b c}, 
\end{equation}
\begin{equation}
\psi_{\rm color}^{f-{\rm type}}(a,b,c)=\sqrt {1\over 24}f^{a b c}.
\end{equation}
After contracting these color wavefunctions
with the short distance potential of Eq. (\ref{m69}),
we get the pair-wise potential for 3-gluon glueball case
\begin{eqnarray}
V_{d,f}(r) &=&-{\lambda \over 2}\Biggl\{
\biggl[{1 \over 4}+{1 \over 3}{\bf S}^2 
+{3 \over 2m^2}({\bf L}\cdot{\bf S}) 
{1 \over r}{\partial \over \partial r} \biggr. \Biggr. \nonumber \\
&& -{1 \over 2m^2}\biggl.
\Bigl(({\bf S}\cdot\nabla)^2-{1 \over 3}{\bf S}^2\nabla^2\Bigr)\biggr]
{e^{-mr} \over r} \nonumber \\
&& +\Biggl. \Bigl(\pm 1-{5 \over 6}{\bf S}^2\Bigr)
    {\pi \over m^2}\delta^3({\bf r})
\Biggr\} +m(1-e^{-\beta mr}),
\label{m92}
\end{eqnarray}
for $d$- and $f$-type, respectively.

\section{Glueball masses and sizes} 

In this paper, we consider only the case of $L=0$,
hence ignore the spin-orbit and tensor terms 
in Eqs. (\ref{m82}) and (\ref{m92}).
%
%
The 2-gluon and 3-gluon glueball systems are discussed separately.

\subsection{Two-gluon Glueballs}

For 2-gluon glueballs with $L=0$, we have only
$J^{PC}=0^{++},2^{++}$ states. The Hamiltonian is 
\begin{equation}
H=2m-{1\over m}\nabla^2+V_{2g},
\end{equation}
where
\begin{eqnarray}
V_{2g}(r) &=& -\lambda\biggl[
\Bigl({1 \over 4}+{1 \over 3}{\bf S}^2\Bigr){e^{-mr} \over r} 
+\Bigl(1-{5 \over 6}{\bf S}^2\Bigr)
 {\pi \over m^2}\delta^3({\bf r})\biggr]\nonumber \\
&& +2m(1-e^{-\beta mr}). 
\end{eqnarray}
We immediately notice one serious problem: 
when $S=0$, one has an attractive $\delta$-function term and
the Hamiltonian is unbounded from below. 
This ``maximum attraction channel" in $0^{++}$
{\it could be related to} 
the gluon condensation that triggers confinement.

In Ref.~\cite{cands},
the $\delta$-function term was treated as a perturbation.
For our study, we propose a physical solution by
smearing the gluon fields,
that is, we replace the $\delta$-function
by the smearing function
\begin{equation}
D(r)={{k^3m^3}\over \pi^{3\over 2}}e^{-k^2m^2r^2}
\label{m38}
\end{equation}
which approaches $\delta^3({\bf r})$ for $k\longrightarrow\infty$.
Using the variational method with trial wavefunction
$\psi (r)\propto e^{-a^2m^2r^2}$, we illustrate in Fig. 2 
the smearing dependence of 2-gluon glueball masses 
for $\lambda=2$ and $\beta=0.3$.

We see from Fig. 2 that
the mass of the $2^{++}$ glueball converges rapidly to $3.2\,m$
for $k > 1$, illustrating good behavior
since the $\delta$-function term is repulsive,
but the $0^{++}$ mass decreases monotonically 
until it becomes negative for $k \gtrsim 3.8$.
To illustrate what is happening when the attractive $\delta$ 
becomes operative for larger $k$ (less smearing),
we plot in Fig. 3 the root-mean-squared radius
$r_{\rm rms}=\sqrt {\langle r^2 \rangle}$ of 
$0^{++}$ and $2^{++}$ glueballs as a function of $k$ 
for the same values of $\lambda$ and $\beta$.
It is clear that, while
the size of $2^{++}$ glueball stabilizes for $k \gtrsim 1$,
the radius for $0^{++}$ glueball drops monotonically
with its mass, which in turn drops monotonically with increase of $k$.

\begin{figure}[b!] 
\centerline{{\epsfxsize3.3 in \epsffile{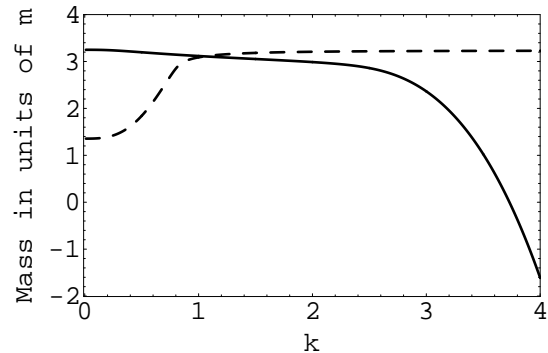}}} 
\smallskip 
\caption{$M_{0^{++}}$ (solid) and $M_{2^{++}}$ (dashed)  
vs. smearing parameter $k$ for $\lambda$=2, $\beta$=0.3. 
In large $k$ limit, one recovers the $\delta$-function potential 
which drives $M_{0^{++}}$ negative..   
}  
\end{figure}

\begin{figure}[t!]  
\centerline{{\epsfxsize3.3 in \epsffile{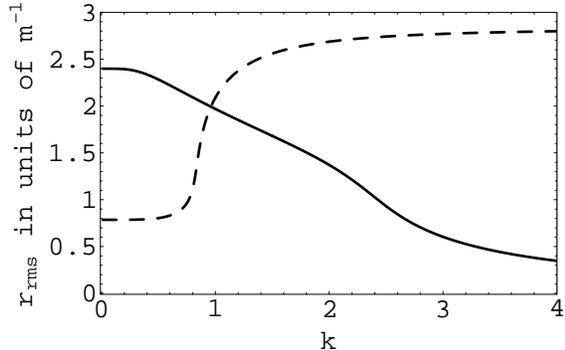}}}  
\smallskip 
\caption{Glueball radius for $0^{++}$ (solid) and $2^{++}$ (dashed)  
glueballs vs. smearing parameter $k$.}  
\end{figure}  

One may be tempted to use Fig. 2 and argue that,
since for $1< k < 3$, both $2^{++}$ and $0^{++}$ glueball masses
are relatively stable, hence they have approximately of the same mass.
However, this is {\it not what is observed} on the lattice
nor as suggested by experiment.
Since we cannot claim to know how to determine 
the value of $k$ in Eq. (\ref{m38}), 
we use the converging experimental and
lattice
 results to fit for $k$.
We take $M_{0^{++}}=1730$ MeV,
and $M_{2^{++}}=2400$ MeV~\cite{Morningstar,Bugg}, 
hence the mass ratio
\begin{equation}
{M_{2^{++}} \over M_{0^{++}}}=1.39.
\label{rat}
\end{equation}
%
Using Eq. (\ref{rat}) to determine $k$,
we find that it depends mainly on $\lambda$, 
but is almost independent of $\beta$.
This is to be expected since the need for smearing comes from
the short distance potential.
With the mass ratio fixed, 
we find a smaller $\lambda$ (and a greater $\beta$ in general)
can accommodate a larger $k$ value, which is reasonable. 
The typical value of $k$ falls in the range of 2.3 to 4.3,
which ensures that the $2^{++}$ mass is stable.

\begin{figure}[b!] 
\centerline{{\epsfxsize3.3 in \epsffile{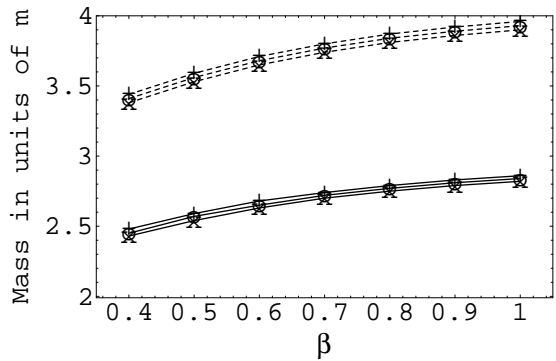}}} 
\smallskip
\caption{$M_{0^{++}}$ and $M_{2^{++}}$ vs. $\beta$ for 
$\lambda =$ 1.5, 2.0 and 2.5 as indicated by 
the symbols ``+", ``o" and ``x", respectively.  
To guide the eye, the points are linked by solid and dashed lines.
The mass ratio is held fixed by Eq. (\ref{rat})
}
\end{figure} 

\begin{figure}[t!]  
\centerline{{\epsfxsize3.3 in \epsffile{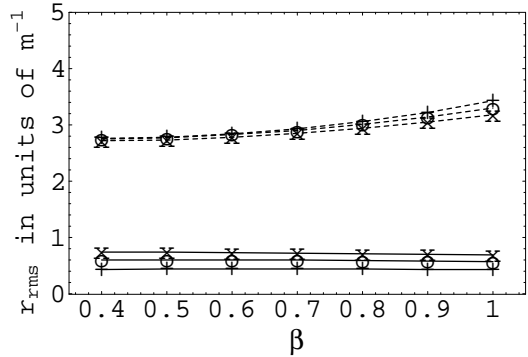}}}  
\caption{Glueball $r_{\rm rms}$ radius for $0^{++}$ and $2^{++}$ glueballs  
vs. $\beta$ for $\lambda =$ 1.5, 2.0 and 2.5 as indicated by the symbols  
``+", ``o" and ``x", respectively, and linked by solid and dashed lines.}  
\end{figure}

With $k$ determined for given $\lambda$ and $\beta$, 
the mass and size of 2-gluon glueballs can be calculated.
Since $M_{0^{++}}$ is lighter than $M_{2^{++}}$
from experiment/lattice input, the $0^{++}$ glueball is rather small in size
compared to the $2^{++}$ glueball.
This is reasonable because of the attractive (repulsive) 
$\delta$ function potential for $0^{++}$ ($2^{++}$).
Except for this smallness in size of $0^{++}$ glueball,
which is dominated by $\lambda$ part of potential,
the $0^{++}$ mass, and the mass and size of the $2^{++}$ glueball are 
all more dependent on $\beta$, the string part of potential.
We plot the masses (in units of $m$) and sizes (in units of $1/m)$ 
vs. $\beta$ for three different values of
$\lambda$ in Figs. 4 and 5, respectively.
These figures illustrate the range of uncertainties
within the model.
We find that the masses increase
with increasing $\beta$ or decreasing 
$\lambda$; however, increasing $\beta$ or decreasing
$\lambda$ will increase the size of $2^{++}$ glueball but decrease the 
size of $0^{++}$ glueball. 
We will give a more detailed discussion on this later.

As mentioned already, the $2^{++}$ glueball mass is 
stable and almost independent of $k$ for $k > 2$, 
which holds for all solvable ($\lambda$, $\beta$) parameter space.
Hence, taking the $2^{++}$ glueball mass 
range of $3.4\,m$ to $3.9\, m$ from Fig. 4
and the lattice result of $M_{2^{++}}$=2400 MeV, 
we estimate the effective gluon mass to be $0.6 - 0.7$ GeV,
which agrees well with 
the gluon mass $\sim$ 0.66 GeV needed~\cite{Consoli,Natale} 
to explain the photon spectrum in radiative
 $J/\psi$ decay.
With gluon mass $m$ determined, 
the typical size of $2^{++}$ and $0^{++}$ glueballs 
can be read off from Fig. 5,
which are in the ranges of 
$0.8 - 1.1$ fm and $0.1-0.2$ fm, respectively,
which is in good agreement with the results \cite{deForcrandLiu}
obtained on a lattice using the source method. For a more 
direct calculation of the $0^{++}$ glueball mass on the lattice,
one would need relatively fine lattice spacings~\cite{Morningstar2}.
It would be interesting to see if our
result of small $0^{++}$ size could be 
further replicated on the lattice.

\subsection{Three-gluon Glueballs}

For 3-gluon glueballs, 
we introduce the center-of-mass and relative coordinates:
${\bf R}=({\bf r_1}+{\bf r_2}+{\bf r_3})/{\sqrt 3}$,  
${\bf r_{12}}=({\bf r_1}-{\bf r_2})/{\sqrt 2}$, and
${\bf r}= {\sqrt {2/3}}({\bf r_1}+{\bf r_2}-2{\bf r_3})$.
Since we consider only pure $s$-states, 
one has the pair potential
\begin{eqnarray}
V_{d,f}(r)&=&-{\lambda\over 2}\biggl[
\Bigl({1 \over 4}+{1 \over 3}{\bf S}^2_{\rm pair}\Bigr){e^{-mr} \over r} 
+\Bigl(\pm 1-{5 \over 6}{\bf S}^2_{\rm pair}\Bigr)\biggr. \nonumber \\
& &\biggl. {\pi \over m^2}\delta^3({\bf r})\biggr] 
+m(1-e^{-\beta mr}),
\end{eqnarray}
where ${\bf S}_{\rm pair}$ is the spin of any pair of gluons
in a 3-gluon glueball system. 
The Hamiltonian for this system is then
\begin{eqnarray}
H_{d,f}&=&3m-{1 \over 2m}\nabla^2_{\bf r}-{1 \over 2m}\nabla^2_{{\bf r}_{12}}\nonumber \\
&+&V_{d,f}(r_{12})+V_{d,f}(r_{23})+V_{d,f}(r_{31}).
\end{eqnarray}

Since the glueball wavefunction must be symmetric with respect to ${\bf
r}_1$, ${\bf r}_2$, and ${\bf r}_3$, the contributions of the three pair
potentials are  the same. Hence the spin and spatial parts of the
glueball wavefunction  are independent and 
\begin{eqnarray}
\sum {\bf S}^2_{\rm pair}&=&({\bf S}_1+{\bf S}_2)^2+({\bf S}_2+{\bf S}_3)^2+
({\bf S}_3+{\bf S}_1)^2\nonumber \\
&=&({\bf S}_1^2+{\bf S}_2^2+{\bf S}_3^2)+({\bf S}_1+{\bf S}_2+
{\bf S}_3)^2.
\end{eqnarray}
The Hamiltonian above can then be simplified as 
\begin{equation}
H_{d,f}=3m-{1 \over 2m}\nabla^2_{\bf r}-{1 \over 2m}\nabla^2_{{\bf r}_{12}}
+V^{T}_{d,f}(r_{12})
\end{equation}
where 
\begin{eqnarray}
V^{T}_{d,f}(r_{12})=&-&{\lambda\over 2}\biggl[ \Bigl({1 \over 4}
+{1 \over 3}(6+{\bf S}^2_{\rm total})\Bigr){e^{-mr_{12}} \over r_{12}}\biggr. \nonumber \\ 
&+& \biggl.  \Bigl(\pm 1-{5 \over 6}(6+{\bf S}^2_{\rm total})\Bigr){\pi \over m^2}
\delta^3({\bf r_{12}})\biggr] \nonumber \\
&+&m(1-e^{-\beta mr_{12}}). 
\end{eqnarray}
where ${\bf S}_{\rm total}$ is now the {\it total spin} of the system,
i.e. $J$.
In Ref.~\cite{Hou}, one introduced an additional quantum number 
called $S_{\rm pair}$ in $J^{PC}(S_{\rm pair})$
which is clearly not adequate,
resulting in a spurious $1^{--}$ state.

\begin{figure}[b!]  
\centerline{{\epsfxsize3.3 in \epsffile{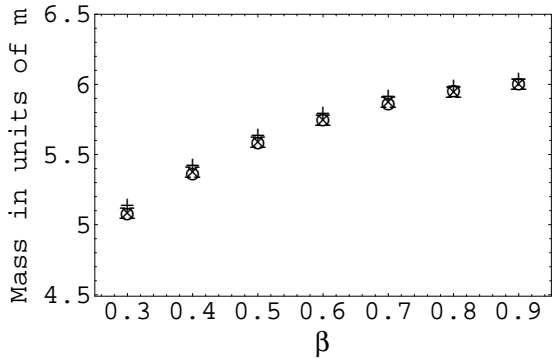}}}  
\smallskip  
\caption{Three-gluon glueball masses  
$M_{0^{-+}}$, $M_{1^{--}}$ and $M_{3^{--}}$ vs. $\beta$,  
as indicated by the symbols ``+", ``o" and ``x", respectively. 
}  
\end{figure}

\begin{figure}[t!]  
\centerline{{\epsfxsize3.3 in \epsffile{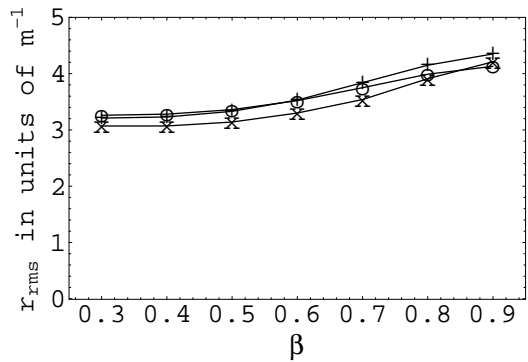}}}  
\smallskip  
\caption{Glueball radius vs. $\beta$ for $0^{-+}$, $1^{--}$ and $3^{--}$   
3-gluon glueballs, as denoted by the symbols ``+", ``o" and ``x", 
respectively. 
To guide the eye, the points are linked by solid lines. 
}  
\end{figure}

There are three possible pure $s$-states, 
with quantum numbers $J^{PC}=0^{-+}$, $1^{--}$, and $3^{--}$.
For these glueballs, all the $\delta$ terms are repulsive,
hence their masses are automatically bounded from below, 
and there is no need to smear the $\delta$ function. 
We estimate the masses and sizes of these glueballs 
using the variational method with the wavefunction
$\psi (r)\propto e^{-a^2m^2({\bf r}^2+{\bf r}^2_{12})}$, which is
symmetric with respect to ${\bf r}_1$, ${\bf r}_2$, and ${\bf r}_3$. 
We find that the masses and sizes are almost $\lambda$ independent,
hence we plot them only vs. $\beta$ in Figs. 6 and 7, respectively.
Note that for the $0^{-+}$ quantum number 
there is also a 2-gluon glueball in $L=1$ state, 
which is expected to have mass lower than the 3-gluon
state discussed here~\cite{cands}.
Mixing between the 2-gluon and 3-gluon states
should lead to level repulsion and raise the $0^{-+}$ 3-gluon
state above the two other states.

From Fig. 6, we find that the masses of the three 
lowest lying 3-gluon glueballs are 
within $0.1\,m$ (or 100 MeV) of each other, 
which holds for any $\beta$ value. 
In other words, they are nearly degenerate, which we will 
discuss further in the next section.
We note that the ratio with $2^{++}$ mass is around 1.53,
almost independent of $\beta$ and $\lambda$,
and agrees with naive constituent counting.
Scaling from the lattice result of
 $M_{2^{++}} = 2000 - 2400$ MeV
the mass range of these glueballs is $3.1-3.7$ GeV,
right in the ballpark of $J/\psi$ and $\psi^\prime$ masses.
Their sizes are only slightly larger than the $2^{++}$ 2-gluon glueball,
and
 fall into the range of $0.9-1.3$ fm.

\section{Discussion and Conclusion} 

We approximated the attractive $\delta$-function by 
a smearing function 
$D(r)$ in Eq. (\ref{m38}). 
This is physically reasonable since 
the gluons in a glueball can not have very large relative momentum
that corresponds to very short distance separation.
Through the variational method, 
besides the constituent terms $2m$ and $3m$, 
one can check four sources for the glueball mass: 
kinetic energy, Yukawa and delta function potential, and string energy. 
It is clear that smaller glueball size implies a greater 
kinetic energy contribution to its mass. 
The Yukawa term is always attractive 
and contributes negatively to the glueball mass.
It is proportional to the adjoint strong coupling constant $\lambda$,
Eq. (\ref{lam}), and becomes more negative for smaller glueball size. 
The delta-function term, 
replaced by the smearing function $D(r)$ of Eq. (\ref{m38}), 
is also proportional to $\lambda$, 
and becomes stronger for smaller glueball size. 
As for string energy term, a bigger glueball can store more energy
hence a larger glueball mass.
A greater string tension $\beta$ also stores more energy in the string.
Knowing about these four sources, we can see how the mass and size
of a glueball comes about. 

Let us first consider masses. 
For the $2^{++}$ glueball, 
the kinetic energy is relatively small due to its rather large size 
compared to the $0^{++}$ glueball. 
On the other hand, there is a cancellation
between $\lambda$-dependent attractive Yukawa potential 
and repulsive delta-function terms.
The resulting value is always negative, 
and will further cancel against the kinetic energy term.
Hence the main contribution to the mass comes from the string energy,
and the stronger the string tension $\beta$, the heavier the mass.
For fixed string tension, a larger $\lambda$ value gives 
a stronger cancellation between 
the remnant of short distance potential and the kinetic energy, 
resulting in a smaller contribution to the glueball mass. 
In other words,
 the $2^{++}$ glueball mass increases 
with
 increasing $\beta$ or decreasing $\lambda$,
as can be seen in Fig. 4.

For the $0^{++}$ glueball, the kinetic 
enengy term is relatively large because of the small size, 
which in turn is brought about by the attractive $\delta$ function,
and there is strong cancellation between these two terms.
Since we have fixed the mass ratio with $2^{++}$, it turns out that
the kinetic energy overcomes the attractive $\delta$ function, 
and subsequently cancels against the attractive Yukawa term. 
The net result could be positive or negative
depending on smaller or larger $\lambda$ value. 
The main contribution still comes from the string term 
but it cannot be as strong as in the case of $2^{++}$ 
because of the small size. 
We stress again that in Fig. 4
we have held the $2^{++}$ and $0^{++}$ ratios fixed according to
Eq. (\ref{rat}).

We turn to the consideration of sizes,
{\it i.e.} understanding Fig. 5. 
We find that the glueball size 
increases with mass for $2^{++}$ glueball.
This is easily understood in terms of a repulsive $\delta$
and the dominance of string energy.
For $0^{++}$ glueball, its size is sensitive to $\lambda$,
which can be understood as coming indirectly from smearing.
As a consequence of imposing Eq. (\ref{rat}), 
the value for $k$ depends mainly on $\lambda$ 
and can become larger for lower $\lambda$.   
For larger $k$, the smearing function $D(r)$ 
approaches the delta function and is more attractive.
Thus, paradoxially, a smaller $\lambda$ (``less attractive") 
is more able to ``pull in" the glueball, 
leading to decrease in size.
However, the size of the $0^{++}$ glueball
is almost $\beta$ independent precisely because of its small size,
hence insensitive to the string energy.
We note that the smallness of the $0^{++}$ glueball
stretches the applicability of our relativistic expansion.
It is therefore amusing that Figs. 4 and 5 are in
rather good agreement with the findings of Ref. \cite{deForcrandLiu}.

For three-gluon glueballs, 
there are three possible pure $S$ states, namely 
$J^{PC}$=$0^{-+}$, $1^{--}$, and $3^{--}$.
For these glueballs, 
all the $\delta$ terms in the potential are repulsive,  
hence no smearing is needed. 
There is a
 $\lambda$-dependent cancellation 
between the Yukawa and delta-function 
terms. 
Interestingly, both terms become stronger for larger total spin (or $J$),
and the cancellation hides the spin effect, 
resulting in
 the masses being nearly degenerate. 
In other words, 
their masses and sizes are almost $\lambda$-independent and 
depend basically on $\beta$, just like $2^{++}$ case. 
It is rather intriguing that the mass difference of the three 
glueballs are within 100 MeV of each other. 
But as we have mentioned earlier,
the $0^{-+}$ state would become heavier via
mixing with the $L = 1$ two-gluon state.

Comparing Figs. 4 and 6,
we note that the 3-gluon glueball masses are 
about 1.5 times larger than $2^{++}$,
largely independent of $\beta$ and $\lambda$. 
Taking $M_{2^{++}}$ between $=2267 \pm 104$ MeV~\cite{Weingarten}
and $2400\pm 25 \pm 120$ MeV~\cite{Morningstar}, 
one finds $M_{1^{--}}$ in the range of $3.5-3.7$ GeV,
close to the $\psi^\prime$ mass of 3686 MeV.
We note that the proximity of the $1^{--}$ glueball to 
$\psi^\prime$ may be called for
from comparsion of $J/\psi$ and $\psi^\prime$
two body hadronic decays~\cite{Suzuki}.
However, if the $f_2(1980)$ state is the $2^{++}$ glueball~\cite{Bugg},
then we find $M_{1^{--}}$ is of order 3 GeV or closer to the $J/\psi$,
where an older proposal~\cite{O} of $1^{--}$ glueball (called $O$)
could be behind the rather sizable strength of 
$J/\psi\to \rho\pi \sim 1\%$.
At the moment, improved action lattice results~\cite{Morningstar} find
$1^{--}$ glueball masses heavier than those discussed here.
We urge further refined, dedicated studies to 
help clarify the phenomenology.
Direct search methods for the lowest lying 3-gluon glueballs
were discussed in Ref.~\cite{Hou},
which should also be brought up to date.

In conclusion, we investigate the bound states of two
and three massive gluons with a Schr\"odinger equation. 
We calculate the short distance potential from one gluon exchange, 
and give arguments for the long distance confining potential. 
We calculate glueball masses and sizes using variational method. 
By considering the effect of smearing of the gluon fields, 
we find that the size of $0^{++}$ glueball could be 
rather small compared with others.
The other glueball masses are stable with respect to such smearing.
Using the converging experimental and lattice results for
$0^{++}$ and $2^{++}$ glueballs, 
we estimate the effective gluon mass to be 0.6 to 0.7 GeV,
in agreement with  phenomenological results.
The typical size of $2^{++}$ and $0^{++}$ are 
of order 1 fm and 0.1 to 0.2 fm, respectively.
This means that to extract the $0^{++}$ glueball size
on a lattice, one would need rather fine lattice spacings. 
It would be of interest to see if our result
would be replicated on the lattice.
For 3-gluon glueballs, their sizes
are also estimated to be in the range of 0.9-1.3 fm,
similar to that of the $2^{++}$ state. 
We find that the three lowest lying 3-gluon
glueballs to be largely degenerate.
In particular, the mass of the $1^{--}$ glueball is 
in the range of 3 to 3.7 GeV, 
which is consistent with arguments from 
phenomenological point of view.

\vskip 0.3cm
\noindent{\bf Acknowledgement}.\ \
This work is supported in part by
the National Science Council of R.O.C.
under Grant NSC-89-2112-M-002-063.



\begin{references}

\bibitem{cornwall}
J.M. Cornwall, Phys. Rev. D {\bf 26}, 1453 (1982).

\bibitem{parisi}
G. Parisi and R. Petronzio, Phys. Lett. {\bf 94B}, 51 (1980).

\bibitem{halzen}
F. Halzen, G.I. Krein and A.A. Natale, Phys. Rev. D~{\bf 47}, 295 (1993);
M.B. Gay Ducati, F. Halzen and A.A. Natale, {\it ibid.}
D {\bf 48}, 2324 (1993);
J.R. Cudell and B.U.~Nguyen, Nucl. Phys. {\bf B420}, 669 (1994).

\bibitem{field}
J.H. Field, Int. J. Mod. Phys. {\bf A9}, 3283 (1994), 
J.P. Liu and W. Wetzel, 
hep-ph/9611250.

\bibitem{Consoli} 
M. Consoli and J.H. Field, Phys. Rev. D {\bf 49}, 1293 (1994);
J. Phys. {\bf G23}, 41 (1997).  

\bibitem{Papavassiliou}
J.R. Forshaw, J. Papavassiliou and C. Parrinello, 
Phys. Rev. D {\bf 59}, 074008 (1999).

\bibitem{Natale}
A. Mihara and A.A. Natale, Phys. Lett. {\bf B482}, 378 (2000).

\bibitem{Bernard}
C. Bernard, Phys. Lett. {\bf 108B}, 431 (1982);
C. Bernard, Nucl. Phys. {\bf B219}, 341 (1983).

\bibitem{amundsen}
P.A. Amundsen and J. Greensite, Phys. Lett. {\bf 173B}, 179 (1986); 
J.E. Mandula and M. Ogilvie, {\it ibid.} {\bf 185B}, 127 (1987); 
R. Gupta {\it et al.}, Phys. Rev. D {\bf 36}, 2813 (1987).

\bibitem{Leinweber}
F.D.R. Bonnet, P.O. Bowman, D.B. Leinweber, A.G.~Williams, 
Phys. Rev. D {\bf 62}, 051501 (2000);
UKQCD Collaboration, D.B.~Leinweber {\it et al.}, 
{\it ibid.} D {\bf 60}, 094507 (1999);
{\it ibid.} D {\bf 58}, 031501 (1998); 
C.~Bernard, C. Parrinello, A.~Soni, {\it ibid.} D {\bf 49}, 1585 (1994);
P. Marenzoni, G.~Martinelli, N. Stella, M. Testa, 
Phys. Lett. {\bf B318}, 511 (1993).

\bibitem{Morningstar}
C.J. Morningstar and M.J. Peardon, Phys. Rev. D {\bf 60}, 034509 (1999).

\bibitem{Weingarten}
A. Vaccarino and D. Weingarten, Phys. Rev. D {\bf 60}, 114501 (1999).

\bibitem{Liu}
C. Liu, hep-lat/0010007;
J. Sexton, A. Vaccarino and  D. Weingarten, 
Phys. Rev. Lett. {\bf 75}, 4563 (1995);
G.S. Bali {\it et al.}, Phys. lett. {\bf B309}, 378 (1993).

\bibitem{Bugg}
D.V. Bugg, M.J. Peardon and B.S. Zou, Phys. Lett. {\bf B486}, 49 (2000).

\bibitem{Anisovich}
A.V. Anisovich {\it et al.}, Phys. Lett. {\bf B471}, 271 (1999) 
and Nucl. Phys. {\bf A662}, 319 (2000); 
D.V. Bugg {\it et al.}, Phys. Lett. {\bf B353}, 378 (1995); 
E760 Collaboration, T.A. Armstrong {\it et al., ibid.}
 {\bf B307}, 394 (1993).

\bibitem{Bugg2} 
D.V. Bugg, L.Y. Dong and B.S. Zou, Phys. Lett. {\bf B458}, 511 (1999);
DM2 Collaboration, D. Bisello {\it et al., ibid.} 
{\bf B192}, 239 (1987) and Phys. Rev. D {\bf 39}, 701 (1989);
MARK-III Collaboration, R.M. Batlrusaitis {\it et al., ibid.}
D {\bf 33}, 1222 (1986).

\bibitem{Barberis}
WA102 Collaboration, D. Barberis {\it et al.}, 
Phys. Lett. {\bf B471}, 440 (2000);
A.V. Anisovich {\it et al., ibid.} {\bf B449}, 145 (1999);
WA91 Collaboration, F. Antinori {\it et al., ibid.} 
{\bf B353}, 589 (1995).


\bibitem{cands}
J.M. Cornwall and A. Soni, Phys. Lett. {\bf 120B}, 431 (1983);

\bibitem{Hou}
W.S. Hou and A. Soni, Phys. Rev. D {\bf 29}, 101 (1984).

\bibitem{deForcrandLiu}
P. de Forcrand and K.F. Liu, Phys. Rev. Lett. {\bf 69}, 245 (1992);
see also T.A. DeGrand, Phys. Rev. D {\bf 36}, 176 (1987).

\bibitem{Morningstar2} 
C. Morningstar and M.J. Peardon,
Nucl. Phys. Proc. Suppl. {\bf 83}, 887 (2000).

\bibitem{Suzuki} M. Suzuki, hep-ph/0006296.

\bibitem{O}
W.S. Hou and A. Soni, Phys. Rev. Lett. {\bf 50}, 569 (1983); 
Ref.~\cite{Hou};
S.J. Brodsky, G.P. Lepage and S.F. Tuan,
Phys. Rev. Lett. {\bf 59}, 621 (1987);
W.S. Hou, Phys. Rev. D {\bf 55}, 6952 (1997).

\end{references}
\end{document}